\documentstyle[aps,preprint]{revtex}
\def\today{\ifcase\month\or
           January\or February\or March\or April\or May\or June\or
           July\or August\or September\or October\or November\or
           December\fi
           \space\number\day, \number\year}

\begin{document}
\draft
\preprint{\vbox{\hfill DOE/ER/40762--028\\
                \null\hfill UMPP \#94--086}}
\title
{Model Independent Extraction of $|V_{\rm bc}|$ Without Heavy Quark Symmetry.}
\author{T. D. Cohen and J. Milana}
\address
{Department of Physics, University of Maryland\\
College Park, Maryland 20742}
\date{February, 1994}
\maketitle
\begin{abstract}
A new method to extract $|V_{\rm bc}|$ is proposed based on a sum--rule
for semileptonic
decays of the $B$ meson.  The method relies on much weaker assumptions
than previous
approaches which are based on heavy--quark symmetry.  This sum--rule
only relies on the
assumption that the virtual $c \overline{c}$ pair content of the $B$
meson can be neglected.
The extraction of the CKM matrix element also requires that the
sum--rule saturates in the
kinematically accessible region.
\end{abstract}
\narrowtext
The CKM quark mixing matrix element $V_{\rm bc}$, between bottom, $b$,
and charm, $c$,
quarks, is
one of the fundamental  parameters of the standard model.   Direct
measurements of this
matrix element at the quark level are not possible because the quarks
are confined in
hadrons.  Thus, to extract $V_{\rm bc}$ from experimental measurements it is
necessary to understand the strong interactions well enough to relate
the observables, which
involve hadrons, to the underlying quark dynamics.   In principle, if
one knew the exact
quantum state for an initial  hadron containing a $b$ quark and an
exact quantum final
state containing a  $c$  quark, one could extract the magnitude of
$V_{\rm bc}$ from a
weak decay from the initial ``$b$'' hadron to the final state.

 Until recently it was believed that until reliable {\it ab initio}
calculations were
available directly from QCD via lattice simulations, one was
unadvoidabley forced to
rely on low energy models for the hadronic matrix elements.  However,  as was
noted
by Nussinov and  Wetzel\cite{NW} and  by Isgur and Wise\cite{WG}, if
the assumptions
underlying heavy quark symmetry are valid, there is a model independent
way to extract
$|V_{\rm bc}|$ from exclusive semi-leptonic weak decays of the
$B$ meson into a $D$ or $D^*$ meson; analogously, semi-leptonic decays of the
$\Lambda_b$ into $\Lambda_c$ could be used.  An alternative method to extract
$V_{\rm bc}$ from semi-leptonic decays   has been proposed by Bjorken, Dunietz
and
Tarron (BDT)\cite{BDT}.   The BDT method also depends on the validity of the
assumptions underlying heavy quark symmetry;   if valid, BDT show that
inclusive $B$ meson or $\Lambda_{b}$ semi-leptonic decays into charmed
states at fixed
three momentum satisfy a sum-rule.
In fact, this sum rule basically just reproduces the spectator model, and hence
by experimentally measuring the semi-leptonic partial width, one
directly extracts $|V_{bc}|$.
In the present letter we observe,  there exists  a sum-rule for semi-leptonic
$B$ or $\Lambda_{b}$ decays similar in spirit to BDT but with much
weaker assumptions.
In particular, the only assumption we need to
make to derive our sum rule  is  that the $\overline{c} c$ content of
the $B$ meson or
$\Lambda_{b}$ can be neglected.  If we make a second assumption, that
our sum rule is
saturated in the kinematically allowable region, we have a model
independent extraction
$|V_{\rm bc}|$ from experimental measurements without relying on heavy
quark symmetry.

Underlying both  the Nussinov-Wetzel-Isgur-Wise (NWIW) method
and the  BDT method are the following assumptions about hadronic states
which carry the
quantum  numbers of a heavy quark:\\
i)  The entire heavy quark content of the state is given by a
single valence heavy quark;\\
ii)  the energy and momentum of the state is carried predominantly
by the heavy quark---$i.e.$ that carried
by the light (anti)quarks and gluons are neglible; and\\
iii) the $b$ quark and $c$ quark may be considered heavy.\\
The meaning of assumption i)  is that one may neglect virtual
$\overline{c}c$ pairs and
$\overline{b}b$ pairs in analyzing the state.     All three of these
assumptions are automatically satisfied in QCD in the formal limit of
$m_b,  m_c \rightarrow \infty$.  The consequences  of these three assumptions
are
profound---they imply the existence of a symmetry for such states:
heavy quark symmetry.
The key physical  point is that if these assumptions are
true, the valence heavy quark acts like a static coulomb color source
in the rest frame of
the hadron.   Thus, in this limit,  the light degrees of freedom ``sees'' the
same color source, independent of the flavor  or spin state of the heavy quark.

The essential point of NWIW method  of extracing $|V_{bc}|$ is that
the transition form factor for  $B$ to $D$ semi-leptonic decays with
zero velocity transfer is
unity since the light degrees of freedom see
the same static color source in both the $B$ and $D$ mesons and hence
the ``wavefunctions''
are identical.   A simple way to think about this is to suppose that
the weak interaction
converting a $b$ quark to a $c$ quark at the same velocity were to happen
suddenly.
The state basically does not change.  It is clear, that this analysis
depends on heavy quark symmetry in an essential way.

The BDT method is based on a key fact:  in the heavy quark  limit, the
various form factors
which in general are independent become related\cite{BDT}.  This
greatly simplifies the analysis
of inclusive semi-leptonic decays.  In analyzing a given exclusive decay that
forms the inclusive sum, the reduction in the number of independent
structures in the transition
form factor is exploited to obtain a simple sum rule which reproduces
the spectator model.
Again, it is clear that this
analysis depends on the assumptions of heavy quark symmetry in an essential
way.

Both of these methods depend critically on the fact that both the
initial state containing a $b$
quark and the final state containing
a $c$ quark satisfy the assumptions of heavy quark symmetry.
While it is probably quite reasonable to assume that the $b$
quark is heavy, the $c$ quark is more problematical---after all
it's mass has been estimated to be  1.3--1.7 GeV which is not
that much larger than the typical hadronic scale of 1 GeV.\cite{comment2}
There are two possible sources of contamination of the extraction
of $V_{bc}$ due to the finiteness of $m_{c}$.  The first is that there
may be nonvanishing $\overline{c}c$ contributions to the initial state
and the second is that
the energy and momentum of the final state may not be dominated by the
valence $c$ quark.
Of these concerns, the second is probably more serious.  The
$\overline{c}c$ contributions in
the initial state might be expected to be particularly small.  In the
first place, the relevant mass
scale in an energy denominator for virtual $\overline{c}c$ pairs is $2
m_c$ rather than
$m_c$; moreover, these contributions may well be dynamically small
independent of the mass since they are Zweig rule violating.
Accordingly, it is of some
importance to see whether one can extract
$|V_{\rm bc}|$ from  experiment based only on the assumption that the
$\overline{c}c$ content
of the $B$ meson or $\Lambda_b$ is neglible but without reliance on the
assumption that the energy and
momentum of the final state is dominated by the $c$ quark.  To make our
assumption concrete:
 we assume that all matrix elements of normal ordered operators
containing any $c$ quark
creation or annihilation operators in a $B$--meson or $\Lambda_b$ can
be neglected.
Here we show that, at least in principle,
one can  make such an extraction from semi-leptonic decays.\cite{others}

The partial width, $\Gamma_n (B \rightarrow l(l) + \overline{\nu}_l
(q-l) + X_c (P_n))$,
for the semi--leptonic decay of a $B$--meson into a particular charmed
state $X_c$ is:
\begin{equation}
\frac{d^{\, 6} \Gamma_n}{d^3l d^3q} = \frac{|M|^2 \delta^4(P_B - q - P_n)}
{8 M_B E (q_0 - E) (2 \pi)^2}
\int \prod_{f=1}^{n} \frac{d^3 f}{(2 \pi)^3 2 E_f}\label{decayn}
\end{equation}
where the phase--space is the usual product over all particles in the
final state $X_c$ and
where  $|M|^2$ is the squared amplitude for the decay,
\begin{equation}
|M|^2 = \frac{G_F^2 |V_{bc}|^2}{2} \{8\, l^{\mu \nu} \}
\langle P_B | J_\mu^{W \dagger}(0) | n \rangle  \langle n | J_\nu^W(0) | P_B
\rangle
\label{amp2}.
\end{equation}
In the above, $E$ is the energy of the lepton, $q_0$ that of the lepton
neutrino pair and
\begin{equation}
l^{\mu \nu} = l^\mu q^\nu + l^\nu q^\mu - 2 l^\mu l^\nu - \frac{1}{2}
q^2 g^{\mu \nu}
+ i \epsilon^{\mu \nu \alpha \beta} q^\alpha l^\beta,
\end{equation}
is the familiar Dirac tensor that results, assuming the neutrino to be
massless, from the
summation over the spins of the lepton and neutrino.
The sum over all states $X_c$ can then be written
\begin{equation}
\frac{d^{\, 6} \Gamma}{d^3l d^3q} \equiv \sum_n \frac{d^{\, 6}
\Gamma_n}{d^3l d^3q} =
\frac{ G_F^2 |V_{bc}|^2 }{2 M_B E (q_0 - E) (2 \pi)^2}  l^{\mu \nu} B_{\mu \nu}
\label{decayS}
\end{equation}
where $B_{\mu \nu}(q,P_B)$ is defined in analogy with deep--inelastic
scattering as the tensor
\begin{equation}
B_{\mu \nu}(q) \equiv \sum_n \delta^4(P_B - q - P_n) \int
\prod_{f=1}^{n} \frac{d^3 f}
{(2 \pi)^3 2 E_f}
\langle P_B | J_\mu^{W \dagger}(0) | n \rangle  \langle n | J_\nu^W(0)
| P_B \rangle.
\label{Btens}
\end{equation}

Consider now the sum over all states with fixed three--momentum $\vec{q}$,
\begin{equation}
\overline{B}_{\mu \nu}^{\, <}(\vec{q}) \equiv \int_0^{M_B-M_D} d q_0 B_{\mu
\nu}(q)
\label{Bbar}
= \sum_n \int d^3 x e^{-i \vec{q} \cdot \vec{x}}
\langle P_B | J_\mu^{W \dagger}(0,\vec{x}) | n \rangle \langle n |
J_\nu^W(0) | P_B \rangle
\label{Btensbarless}
\end{equation}
in which the upper limit of the integral has been evaluated in the rest
frame of the $B$--meson
and is the difference in the masses of the $B$ and $D$
mesons.  It is the maximum energy the lepton--neutrino pair can
carry--off as evaluated in this
frame.  Note that we have translated the operator $J_\mu^W$ and used
the standard integral
representation for the $\delta$--function to convert Eq. (\ref{Btens})
into the above
Fourier--transform.  The superscript in our nomenclature is motivated
by the fact that the
definition of $\overline{B}_{\mu \nu}^{\, <}$ involves a sum over all
charmed states $X_c$ with energy (in the rest of the $B$--meson) $P_n^0
< M_B$.  In a
complete sum over states one would also need to include states $X'_c$ such that
$P_n^0 > M_B$.
If however the kinematically available states saturate the summation, then
\begin{equation}
\overline{B}_{\mu \nu}^{\, <}(\vec{q}) \approx \overline{B}_{\mu
\nu}(\vec{q}) \equiv
 \int d^3 x e^{-i \vec{q} \cdot \vec{x}}
\langle P_B | J_\mu^{W\dagger}(0,\vec{x}) J_\nu^W(0) | P_B
\rangle.\label{Btensbar}
\end{equation}
It is $\overline{B}_{\mu \nu}(\vec{q})$ that obeys a sum rule which we
will now derive using
the mildest assumptions concerning the ground state of the $B$--meson.
The issue whether
Eq. (\ref{Btensbar}) is itself a good approximation, $i.e.$ whether the
sum over states saturates,
is ultimately a question that must be determined by experiment.
We will return to this issue after we derive the sum--rule that
$\overline{B}_{\mu \nu}(\vec{q})$ obeys.

We start by using a standard trick of many--body physics\cite{FetWal}
to convert the
equal--time product of fields in Eq. (\ref{Btensbar}) into a time ordered
product:
\begin{equation}
\overline{B}_{\mu \nu}(\vec{q}) = \lim_{t \rightarrow 0^+}
\int d^3 x e^{-i \vec{q} \cdot \vec{x}}
\langle P_B | T( J_\mu^{W\dagger}(t,\vec{x}) J_\nu^W(0) ) | P_B
\rangle.\label{Btime}
\end{equation}
The relevant weak currents are
\begin{equation}
J_\nu^W(x) = : \overline{c}(x) \Gamma_\nu b(x) : \label{current}
\end{equation}
where $c(x)$ and $b(x)$ are the charm and bottom quark field operators
respectively and
\begin{equation}
\Gamma_\nu = \gamma_\nu (1 - \gamma_5 ).\label{projector}
\end{equation}
Applying Wick's theorems to the time--ordered product one obtains that
\begin{eqnarray}
\langle P_B | T( J_\mu^{W \dagger}(t,\vec{x}) &J_\nu^W(0)& ) | P_B \rangle =
\langle P_B | : \overline{b}_\alpha(t,\vec{x}) \Gamma_{\mu , \alpha
\beta} c_\beta(t,\vec{x})
\overline{c}_{\beta^\prime}(0) \Gamma_{\nu , \beta^\prime \alpha^\prime}
b_{\alpha^\prime}(0) : | P_B \rangle \nonumber\\
&-& \langle 0 | T( b_{\alpha^\prime}(0) \overline{b}_\alpha(t,\vec{x})
) | 0  \rangle
\Gamma_{\mu , \alpha \beta} \Gamma_{\nu , \beta^\prime \alpha^\prime}
\langle P_B | : c_\beta(t,\vec{x}) \overline{c}_{\beta^\prime}(0) : |
P_B \rangle\nonumber\\
&+& \langle 0 | T( c_\beta(t,\vec{x}) \overline{c}_{\beta^\prime}(0)) | 0
\rangle
\Gamma_{\mu , \alpha \beta} \Gamma_{\nu , \beta^\prime \alpha^\prime}
\langle P_B | : \overline{b}_\alpha(t,\vec{x}) b_{\alpha^\prime}(0) : |
P_B \rangle \nonumber\\
&-& \langle 0 | T( b_{\alpha^\prime}(0) \overline{b}_\alpha(t,\vec{x})
) | 0  \rangle
\Gamma_{\mu , \alpha \beta} \Gamma_{\nu , \beta^\prime \alpha^\prime}
\langle 0 | T( c_\beta(t,\vec{x}) \overline{c}_{\beta^\prime}(0)) | 0 \rangle
\langle P_B | P_B \rangle,
\label{Wick}
\end{eqnarray}
where $\alpha$ and $\beta$ explicitly label the components in Dirac
space.  The last term
 is of course a disconnected diagram and is ignorable for the case at
hand.\cite{comment}

We now impose our one physical assumption that the $B$--meson has neglible
virtual
$c \overline{c}$ pairs.  This then
eliminates the first two terms on the r.h.s. in Eq. (\ref{Wick}), leaving that
\begin{equation}
\langle P_B | T( J_\mu^{ W \dagger}(t,\vec{x}) J_\nu^W(0) ) | P_B \rangle
\approx
i S^c_{\beta \beta^\prime} (t,\vec{x}) \Gamma_{\mu , \alpha \beta}
\Gamma_{\nu , \beta^\prime \alpha^\prime}
\langle P_B | : \overline{b}_\alpha(t,\vec{x}) b(0)_{\alpha^\prime} : | P_B
\rangle
\label{reduce}
\end{equation}
where $S^c$ is the standard free--space causal (Feynman) propagator for
a charm quark:
\begin{equation}
S^{c} (t,\vec{x}) = \int \frac{(\not\!p + m_c)}{p^2 - m_c^2 + i \epsilon}
\frac{d^4 \, p}{(2 \pi)^4} e^{-i p \cdot x}.
\label{Feyprop}
\end{equation}
Consider now the combination\cite{combination}
\begin{equation}
\overline{B}_E(\vec{q}) \equiv 2 \overline{B}_{0 0}(\vec{q}) -
\overline{B}_{\mu \mu}(\vec{q}).
\label{defEuc}
\end{equation}
{}From Equations (\ref{projector}) and (\ref{Feyprop}) the Dirac algebra
for $\overline{B}_E$
becomes simply $8 p_0 \gamma_0 (1 - \gamma_5)$, and
\begin{eqnarray}
\overline{B}_E (\vec{q}) = \lim_{t \rightarrow 0^+} i 8 \int d^3 x
e^{-i \vec{q} \cdot \vec{x}} & &\langle P_B | : b{^\dagger}(t,\vec{x})
(1 - \gamma_5)  b(0) : | P_B \rangle\nonumber\\
& &\int \frac{p_0 d^4 p}{(2 \pi)^4}\frac{ e^{i \vec{p} \cdot \vec{x} -i p_0
x_0}}
{(p_0 - E_p + i\epsilon) (p_0 + E_p - i\epsilon)}.
\label{Beuclid}
\end{eqnarray}
Performing the integral over $p_0$ by closing on the lower--half plane
(since $t \rightarrow 0^+$)
one obtains that
\begin{eqnarray}
\overline{B}_E (\vec{q}) &=& \lim_{t \rightarrow 0^+} 4 \int d^3 x
e^{-i \vec{q} \cdot \vec{x}}
\langle P_B | : b{^\dagger}(t,\vec{x}) (1 - \gamma_5)  b(0) : | P_B \rangle
\int \frac{d^3 \, \vec{p}}{(2 \pi)^3} e^{i \vec{p} \cdot \vec{x}}\nonumber\\
&=& \lim_{t \rightarrow 0^+} 4 \int d^3 x e^{-i \vec{q} \cdot \vec{x}}
\delta^3(\vec{x})
\langle P_B | : b{^\dagger}(t,\vec{x}) (1 - \gamma_5)  b(0) : | P_B
\rangle\nonumber\\
&=& 4 \langle P_B | : b{^\dagger}(0) (1 - \gamma_5)  b(0) : | P_B \rangle = 8
M_B.
\label{sumResult}
\end{eqnarray}
This then is our sum rule.

Since our result Eq. (\ref{sumResult}) is a simple scalar, it is
Lorentz invariant although the
combination entering the definition of $\overline{B}_E(\vec{q})$ is not
manifestly so.
The corrections to Eq. (\ref{sumResult}) however are in general not
Lorentz invariant.
 However the Lorentz invariance of our result also makes a comparison
with its evaluation
in various frames informative.

As we already mentioned, the experimental verification of Eq.
(\ref{sumResult}) requires
that the sum over states, Eq. (\ref{Btensbar}), is saturated by the
kinematically available
states for $B$--decays.  Itis an issue that can be addressed
experimentally as to whether this assumption holds.  If for fixed
$\vec{q}$, the sum over states entering $B_E(\vec{q})$ is found to
saturate experimentally with
little incremental change as the hadronic final states approaching
$M_B$, then the sum rule
is presumably useful.  We will conclude momentarily with how
such a result can then be used to determine the CKM matrix element, $|V_{bc}|$,
but first some remarks on kinematics and the relation between our
result and that of
Bjorken $et. al.$ \cite{BDT}.

As is well known, the most general hadronic tensor $B_{\mu \nu}(q,P_B)$
in Eq. (\ref{Btens})
can be expanded in terms of six scalar functions:
\begin{eqnarray}
B_{\mu \nu}(q,P_B) &=& B_1 (q,P_B) P_{B \mu} P_{B \nu} + B_2 (q,P_B)
M^2_B g_{\mu \nu}
+ B_3 (q,P_B) (P_{B \mu} q_\nu + q_\mu P_{B \nu})\nonumber\\
&+& B_4 (q,P_B) (P_{B \mu} q_\nu - q_\mu P_{B \nu})
+ B_5 (q,P_B) q_\mu q_\nu + B_6 (q,P_B) \epsilon_{\mu \nu \alpha \beta}
P_{B \alpha} q_\beta.
\label{expansion}
\end{eqnarray}
Note that due to the structure of the lepton's tensor, $l^{\mu \nu}$,
$B_4 (q,P_B)$ never
contributes tothe decay of the $B$--meson.  It would thus be disastrous
if our sum rule
required knowledge of this
function.  Fortunately it does not, as the combination
\begin{eqnarray}
2 B_{0 0}(q,P_B) - B_{\mu \mu}(q,P_B) &=& M_B^2 ( B_1(q,P_B) -
B_2(q,P_B) )\nonumber\\
&+& 2 M_B q_0 B_3(q,P_B) + B_5(q,P_B) (2 q^2_0 - q^2).\label{combination}
\end{eqnarray}
We see therefore that our sum rule requires the determination
(in a Rosenbluth--like fashion) of four
functions for each decay.  Also note from the contraction with $l^{\mu \nu}$,
the
contributions of $B_3(q,P_B)$ and $B_5(q,P_B)$ to a decay rate go like $m_l^2$
and
$m_l^4$ respectively, where $m_l$
is the mass of the outgoing lepton.  Hence we expect that muonic decays
will minimally be
needed for determining $B_3(q,P_B)$, while $\tau$ decay modes will most
likely be needed
for $B_5(q,P_B)$.

The sum--rule of BDT is a special case of ours when using the
``spectator model'' assumptions that
(i) the hadronic matrix elements are simply proportional to the free
$b \rightarrow c$ quark transition, and
(ii) that the momentum of the hadronic final--state, $P_n$, is
dominated by that of the
charm quark.  With these two assumption, the hadronic matrix element
$B_{\mu \nu}(q,P_B)$
is then given by:
\begin{equation}
B_{\mu \nu}(q,P_B) = \frac{2 M_B F^2(q,P_B)}{E_b E_c}  \left\{ 2 P_{B \mu}
P_{B \nu} - (P_{B \mu} q_\nu
+ P_{B \nu} q_\mu) - g_{\mu \nu} (M_B^2 - P_B \cdot q) -
i \epsilon_{\mu \nu \alpha \beta}
P_{B \alpha} q_\beta \right\}\label{hadtensBDT}
\end{equation}
where $F^2$ is a single structure function, $E_b$ and $E_c$ are the energies of
the
$b$ ($P_B$) and $c$ ($P_n$) quarks respectively and are included so
that $F^2$ matches
smoothly onto the zero--momentum recoil limit of heavy--quark
symmetry\cite{NW},
\cite{WG}, $F^2(\vec{q}=0)=1$.
Note that $B_{\mu \nu}(q,P_B)$ involves a sum over the spins of the charm quark
and
an average over that of the bottom.  Matching now the terms of Eq.
(\ref{combination})
and Eq. (\ref{hadtensBDT}) one obtains that
\begin{equation}
2 B_{0 0}(q,P_B) - B_{\mu \mu}(q,P_B) = \frac{ 8 M_B F^2(q,P_B)}{E_b E_c}
(M_B^2 - P_B \cdot q) = 8 M_B F^2.\label{step}
\end{equation}
Inserting Eq. (\ref{step}) into the sum rule, Eq. (\ref{sumResult}), one
obtains the result of BDT:
\begin{equation}
\int d\, q_0 F^2 = 1.\label{BDTresult}
\end{equation}

The connection of this sum rule to the NWIW method is equally obvious.
If all of the assumptions underlying heavy quark symmetry
are satisfied then our sum rule, when evaluated for $\vec{q}=0$ (in the
$B$ meson rest frame)
will be saturated by two states, the
$D$ and $D^*$.  We should stress, however, that the sum-rule methods of BDT and
the
present paper have one advantage over the NWIW approach.  In NWIW,
$|V_{\rm bc}|$ is determined
from a single kinematical point; this does not allow one to test the
validity of the assumptions
underlying the method.  In contrast, the sum-rule methods  allow for
independent extractions of
$|V_{\rm bc}|$ at any $\vec{q}$ (for which the sum   rule saturates).
Thus, the consistency of the
extracted $|V_{\rm bc}|$ for various values of $\vec{q}$ serves as a
test of  the underlying assumptions.

It is also worth noting that interesting physics can be extracted
from the sum rule, independent of the extraction of $|V_{\rm bc}|$.
Comparisons of $|V_{\rm bc}|$ extracted by the present method and
via NWIW and BDT methods gives a test of the validity of the assumptions of
heavy quark symmetry which is certainly interesting in its own right.
Moreover, the weaker assumption of the present sum rule, that the
virtual $\overline{c}c$
pairs are neglible in a $B$ meson, can also be tested by checking  the
consistency of the
extracted $|V_{\rm bc}|$ for various values of $\vec{q}$.

The potential usefulness of this sum-rule to extract $|V_{\rm bc}|$
should be clear.
In order to make such an extraction one must measure the
energy and momentum of both the final lepton and of {\it all} of the
final hadrons.
Simple kinematics allows one to do a  Rosenbluth-like seperation which
experimentally
gives $|V_{\rm bc}|^2 B_{\mu \nu}(q_0, \vec{q})$.   Implementing the
sum rule allows an
extraction of $|V_{\rm bc}|$.  As noted above this extraction may not
be easy technically
since the extraction of $B_{\mu \nu}(q_0, \vec{q})$
contains pieces proportional to the square of the lepton mass,
presumably requiring studies
in which the final lepton is a $\tau$.
Whether or not such an extraction is possible at machines available
either presently or in the near future requires further study.
Clearly, this method is experimentally more difficult than either the
NWIW or BDT methods.  On the other hand, the theoretical uncertainities
associated
with the extracted $|V_{\rm bc}|$ will be greatly minimized.

This work was supported in part by the U.S. Department of Energy under
grant No. DE-FG05-93ER-40762 and NSF grant PHY-9058487.


\begin{references}
\bibitem{comment2} See also \cite{eeprod} and \cite{systematics} for
other cases that
large violations to naive predictions using heavy quark symmetry are
expected in the charm system.
\bibitem{comment} Strictly, we have a sum--rule for
$\langle P_B | J_\mu^{W\dagger}(0,\vec{x}) J_\nu^W(0) | P_B \rangle - \langle 0
| J_\mu^{W\dagger}(0,\vec{x}) J_\nu^W(0) | 0 \rangle$ where $ | 0
\rangle$ is the perturbative vacuum.
We note however that in the kinematically accessible region, there is no
contribution from the perturbative piece.
\end{references}
\end{document}